\documentclass{article}

\usepackage{amssymb,amsmath,epsf}
\usepackage{pictexwd,dcpic}
\usepackage{wasysym}
\begin{document}

\title{On the Geometric Quantization of Canonical Gravity}         
\author{Vasudev Shyam\\ \\Centre for Fundamental Research and Creative Education,\\
Bangalore, India}        
\date{\today}          
\maketitle
\begin{abstract}
One of the hardest problems to tackle in the dynamics of canonical approaches to quantum gravity is that of the Hamiltonian constraint. We investigate said problem in the context of formal geometric quantization. We study the implications of the non uniqueness in the choice of the vector field which satisfies the presymplectic equation for the Hamiltonian constraint, and study the implication of the same in the quantization of the theory. Our aim is to show that this non uniqueness in the choice of said vector field, which really stems from refoliation invariance leads to a very ambiguous notion of quantum evolution. We then investigate the case of a theory where the problem of the Hamiltonian constraint has been dealt with at the classical level, namely Shape Dynamics, and attempt to derive a time dependent Schrodinger equation for the quantum dynamics of this theory.
\end{abstract}
\section{Introduction}       
The process of geometric quantization is one through which the quantum Hilbert space is attained from the classical phase space by means of pre-quantization wherein one constructs a Hermitian line bundle equipped with a connection whose curvature equals the symplectic form of the base space here being the classical phase space of the system under consideration. This is followed by finding a suitable polarization of this pre-quantum line bundle in order to remove `half' the degrees of freedom. Finally, one arrives at the quantum Hilbert space by means of Metaplectic correction. This strategy has been applied to constrained systems too, although, there are some additional steps that are involved in the aforementioned process. In this paper we investigate the implications of applying this procedure to canonical theories of general relativity, where, the Hamiltonian is but a sum of constraints and one of these constraints, namely the scalar Hamiltonian constraint seems to be the source of many of the problems in the dynamics of the quantum theory. In the following section, we shall describe the classical phase space of canonical ADM gravity.
\section{The Classical Phase Space of Canonical Gravity}
In classical ADM general relativity, the configuration space is that of Riemannian three geometries, and the phase space is the cotangent bundle of the same. In this paper we denote it as $\Gamma$. Here, as the system is constrained, the geometry of $\Gamma$ is said to be presymplectic which implies that the counterpart of the symplectic form on the phase space is only weakly non degenerate i.e. it is degenerate everywhere except on the surfaces wherein the constraints vanish (see \cite{gnh}). The Hamiltonian is given by
\begin{equation}
 \mathcal{H}=\int d^{3}x (\frac{N}{\sqrt{\gamma}} G^{abcd}\pi_{ab}\pi_{cd} -N\sqrt{\gamma}R-2\xi^{a}\nabla_{a}\pi^{ab})
\end{equation}
We attain the constraint hypersurface by imposing
\begin{equation}
\frac{\delta \mathcal{H}}{\delta N}|_{\tilde{\Gamma}}=\Phi^{0}[z^{I}]
\end{equation}
\begin{equation}
\frac{\delta \mathcal{H}}{\delta \xi}|_{\tilde{\Gamma}}=\Phi^{1}[z^{I}]
\end{equation}
In totality
$$\Phi^{J}[z^{I}]=0,$$
Here
\[z^{I}=\left(\begin{array}{cc}\pi^{ab}\\ \gamma_{ab}\end{array}\right)\],
 On this hypersurface, there exists the presymplectic form
\begin{equation}\Omega|_{\tilde{\Gamma}}=\int_{\Sigma}\textrm{d}^{3}x \textrm{d}_{\delta}\pi_{ab}\land \textrm{d}_{\delta}\gamma^{ab} .\end{equation}
Here, $d_{\delta}$ is the functional exterior derivative.  We now use the fact that any symplectic vector field on the constraint hypersurface will be locally Hamiltonian for it's flow preserves $\Omega$ i.e. 
$$\mathcal{L}_{X}\Omega|_{\tilde{\Gamma}}=0$$
$$=> (\iota_{X}d_{\delta}\Omega+d_{\delta}\iota_{X}\Omega)|_{\tilde{\Gamma}}=0$$
$$=>(d_{\delta}\iota_{X}\Omega)|_{\tilde{\Gamma}}=0$$
$$=>(\iota_{X}\Omega)|_{\tilde{\Gamma}}=d \mathcal{H}$$
$$i_{*}X=X_{\mathcal{H}}.$$
Here $i$ is the inclusion map from $\tilde{\Gamma}$ to $\Gamma$. From the above calculation we obtain the (locally) Hamiltonian vector field
$$X_{\mathcal{H}}=\left[ 2\frac{N}{\sqrt{\gamma}}\left(\pi_{ab}-\frac{1}{2}\gamma_{ab} tr\pi \right)+ \mathcal{L}_{\xi^{a}}\gamma_{ab}\right]\frac{\delta}{\delta \gamma^{ab}}-$$
$$[ N\sqrt{\gamma}\left(R^{ab}-\frac{1}{2}\gamma^{ab}R \right) - \frac{N \gamma^{ab}}{2\sqrt{\gamma}}\left(\pi_{ab}\pi^{ab}-\frac{1}{2}tr\pi^{2}\right)+\frac{2N}{\sqrt{\gamma}}\left(\pi^{ac}\pi^{b}_{c}-\frac{1}{2}\pi^{ab} tr\pi \right)+$$
$$\sqrt{\gamma}(\nabla^{a}\nabla^{b}N-\gamma^{ab}\nabla^{2}N) + \mathcal{L}_{\xi^{a}}\pi^{ab} ]\frac{\delta}{\delta \pi_{ab}} .$$
Thus, the presymplectic equation is
\begin{equation}
(X_{\mathcal{H}})^{\flat}|_{\tilde{\Gamma}}=0\end{equation}
Here, the map $\flat:T\Gamma\rightarrow T^{*}\Gamma$ is surjective only when restricted to $\tilde{\Gamma}$, in accordance with the Gotay-Nester presymplectic algorithm. It's action is defined as $$\forall Z\in T\Gamma, (Z)^{\flat}=\Omega(Z)$$ on $\tilde{\Gamma}$.
The total Hamiltonian vector field splits up into a vector field that generates evolution and an infinitesimal generator of the diffeomorhism group action on phase space, given by
$$
\mathcal{E}_{H(N)}|_{\tilde{\Gamma}}= 2\frac{N}{\sqrt{\gamma}}\left(\pi_{ab}-\frac{1}{2}\gamma_{ab} tr\pi \right)\frac{\delta}{\delta \gamma^{ab}}-
$$
$$
[ N\sqrt{\gamma}\left(R^{ab}-\frac{1}{2}\gamma^{ab}R \right) - \frac{N \gamma^{ab}}{2\sqrt{\gamma}}\left(\pi_{ab}\pi^{ab}-\frac{1}{2}tr\pi^{2}\right)+\frac{2N}{\sqrt{\gamma}}\left(\pi^{ac}\pi^{b}_{c}-\frac{1}{2}\pi^{ab} tr\pi \right)+
 $$ $$
\sqrt{\gamma}(\nabla^{a}\nabla^{b}N-\gamma^{ab}\nabla^{2}N)]\frac{\delta}{\delta \pi_{ab}},
$$
and
$$
\mathcal{G}_{H_{a}(\xi^{a})}|_{\tilde{\Gamma}}=\mathcal{L}_{\xi^{a}}\gamma_{ab}\frac{\delta}{\delta \gamma^{ab}}+\mathcal{L}_{\xi^{a}}\pi^{ab}\frac{\delta}{\delta \pi_{ab}},
$$
respectively. It can be shown that the constraint corresponding to the latter (i.e. the diffeomorphism constraint) is an equivariant moment map and a symplectic reduction by it is possible. 
\section{Geometric Quantization and Geometrodynamics}
In this section we shall deal with the problems related to the formal geometric quantization of geometrodynamics. In particular we shall investigate the problems with trying to quantize the scalar Hamiltonian constraint to yield the Wheeler--Dewitt equation. First, we do the pre-quantization. The pre-quantization of $(\tilde{\Gamma},\Omega)$ is a line bundle $(L,\mathcal{D}_{\Theta})$; $\pi: L\rightarrow \tilde{\Gamma}$ where
$$\mathcal{D}_{\Theta}\mathcal{D}_{\Theta}=\frac{1}{h}\Omega$$
That is, the curvature of the line bundle equals the symplectic form, upto a multiple of 1 over the Planck's constant.
Consequently, the connection $\mathcal{D}_{\theta}$ is given by:
$$\mathcal{D}_{\Theta}=\textrm{d}_{\delta}-\frac{1}{\hbar}\Theta,$$
where $\textrm{d}_{\delta}\Theta=\Omega$.
 We choose a real polarization $P$ which is an involutive distribution and satisfies $P^{\perp}=P$. In this case $$P=\textrm{span}\left\{\frac{\delta}{\delta \pi_{ab}}\right\}$$
$P$ polarized sections $\Psi$ are those which satisfy
$$\mathcal{D}_{\Theta}(X)\Psi=0  ;\forall X\in P.$$
Finally, in order to attain a Hilbert space, one needs to associate with the above structure an inner product, which shall be left arbitrary in our present discussion for it will not have bearing on the analysis that shall follow. Before we begin our discussion about the quantization of the scalar constraint, we shall first define the flow of it's Hamiltonian vector field. The flow of the evolutionary vector field is given by the solution to the Cauchy problem 
$$f^{0}_{H(N)}(z^{I})=z^{I}$$
$$\frac{\textrm{d}}{\textrm{d}\lambda}f^{\lambda}_{H(N)}(z^{I})=\mathcal{E}_{H(N)}f^{\lambda}_{H(N)}(z^{I})$$
on $\tilde{\Gamma}$. The formal solution to the above is given by
\begin{equation}f^{\lambda}_{H(N)}=\sum_{n}^{\infty}\frac{\lambda^{n}}{n!}\mathcal{E}^{n}_{H(N)}.\end{equation}
Note that 
$$f^{\lambda}_{H(N)}(F(z^{I}))=F(f^{\lambda}_{H(N)}(z^{I}))=F(z^{I};\lambda).$$
The quantized Hamiltonian constraint most definitely changes the polarization on action on Hilbert space. This can be characterized as follows:
As $\hat{H}(N)$ moves $\Psi$ out of $P$, the evolved state
$$\Psi_{\lambda}=\hat{f}^{\lambda}_{H(N)}\Psi$$
is polarized with respect to $P_{\lambda}$ which is the pull back polarization
$$P_{\lambda}=(f^{\lambda}_{H(N)})^{*}P$$
Now, it is straightforward to define the quantum operator corresponding to $H(N)$ as
\begin{equation}\mathcal{Q}(H(N))\Psi=-i\hbar\frac{\textrm{d}}{\textrm{d}\lambda}(\Pi_{\lambda}\hat{f}^{\lambda}_{H(N)}\Psi)|_{\lambda=0}\end{equation}
Here $\Pi_{\lambda}$ is the projection operator from the evolved to the $\lambda=0$ Hilbert space (corresponding with the projection from $P_{\lambda}$ to $P$). It should be noted that we are not going to the full extent of geometric quantization of functions that do not preserve the polarization wherein one would have to deal with the BKS kernel for the quantization of this operator, we simply use this naive form of quantization in order to exhibit the problems associated to quantizing such a constraint. Under the assumption that the scalar constraint generates physical evolution as opposed to gauge motion and the lack of a true Hamiltonian imply that there is nothing holy about the form of the equations of motion of this theory. In order to make this statement more precise, we note that, in general, the constraint sub-manifold that the Gotay Nester algorithm yields need only satisfy the equation
\begin{equation}\langle T\tilde{\Gamma}^{\perp}|\textrm{d}\mathcal{H}\rangle=0\end{equation}
and, 
\begin{equation}(\iota_{X_{\mathcal{H}}}\Omega=\textrm{d}\mathcal{H})|_{\tilde{\Gamma}}\end{equation}
$T\tilde{\Gamma}^{\perp}$ is the symplectic orthogonal of $T\tilde{\Gamma}$. This implies that the dynamical orbit of the system on phase space would correspond to the integral surfaces of the Hamiltonian vector field on the final constraint submanifold, but in the theory we have here, the absence of the true Hamiltonian means that the previous equation would not have a non zero R.H.S even on $\tilde{\Gamma}$. Now, for a general Hamiltonian system constrained (but not totally constrained) or otherwise, the dynamics of the system is uniquely determined by it's deterministic trajectory on phase space (or on the reduced phase space in the case of a constrained theory), and this trajectory is but the integral curve of the Hamiltonian vector field. But, in this theory, due to re-foliation invariance, we see the criterion for said vector field to satisfy the presymplectic equation is too large, and so the flows are not unique, and the non uniqueness in their choice reflects re-foliation invariance. So it would be very ambiguous indeed to try and quantize the Hamiltonian constraint with the above technique, also, the fact that there is one such constraint per space point contributed greatly to this ambiguity in dynamics we profess here. In order to demonstrate this, let us consider a pair of curves on the constraint surface $\tilde{\Gamma}$, and let them be identical (i.e. congruent) upto a phase space three geometry $z_{0}^{I}$ where after they differ only in the choice of lapse, i.e. they would correspond to say, $f^{\lambda}_{H(N')}$ and $f^{\lambda}_{H(N'')}$ and of course, they would just correspond to two different choices in foliating space time, now, if we had a point in $z_{0}^{I}$, there would be two equilocal points in $z_{0}^{I'}$, $z_{0}^{I''}$ associated to two distinct flows (to see more detailed arguments along this line see \cite{kthb}). As it is seen that as the Hamiltonian constraint changes the polarization it acts on, and the way we deal with it is through pulling back the polarization with the flow corresponding to it's vector field, our current discussion on the indeterminism associated with said flows due to re-foliation invariance tells us why it would be problematic to attain a unique $\Pi_{\lambda}$. Also reducing the phase space of canonical gravity is not feasible as quotienting by the action of the scalar constraint would be the same as identifying the past and the future of the system. We shall look now to a more prudent means of solving this problem, that is, the theory of Shape Dynamics. 
\section{The Case for Shape Dynamics}
In this section we shall discuss the theory of Shape Dynamics which is a theory that shares the phase space of ADM gravity but possesses spatial conformal symmetry and the Hamiltonian constraint no longer persists in this theory as it possesses a true Hamiltonian. We shall not delve into the derivation of the theory from the linking theory construction, but we shall present the symplectic reduction of the theory by it's associated symmetries. This theory is ,strictly speaking, not one of geometrodynamics, i.e. it is not a theory that comes out of gauge fixing the ADM phase space. But, a transition can be made between this theory and ADM in CMC (constant mean curvature) gauge, and both ADM gravity and Shape Dynamics are different limits of a larger Linking theory. For further details regarding said details, we refer the reader to (\cite{hk1})
\subsection{Reduction of the Classical Theory}
The Hamiltonian of Shape Dynamics is given by
\begin{equation}\mathcal{H}_{SD}=\int_{\Sigma}(\sqrt{\gamma}e^{6\phi[\gamma,\pi,\tau]}+\pi^{ab}(\mathcal{L}_{\xi^{a}}\gamma_{ab})+\rho\textrm{tr}\pi).\end{equation}
Here $e^{6\phi}$ is the conformal factor. This solves the Lichnerowicz York equation:
\begin{equation}-8\nabla^{2}e^{\phi}+Re^{\phi}-\frac{\pi_{ab}\pi^{ab}e^{-7\phi}}{|\gamma|}+\frac{3}{8}\tau^{2}e^{8\phi}=0\end{equation}
which always has a unique solution.
The physical Hamiltonian is given by
\begin{equation}H_{SD}=\int_{\Sigma}\sqrt{\gamma}e^{6\phi[\gamma,\pi]}\end{equation}
and, the conformal and diffeomorphism constraints are given by
\begin{equation}\mathcal{C}(\rho)=\int_{\Sigma}\rho \textrm{tr}\pi,\end{equation}
\begin{equation}H_{a}(\xi^{a})=\int_{\Sigma}\pi^{ab}(\mathcal{L}_{\xi^{a}}\gamma_{ab}),\end{equation}
respectively. Before moving on to the reduction, we shall first define the group of diffeomorphisms that we are considering. It is the proper subgroup of the Diffeomorphism group where the group action fixes a preferred point $\infty \in \Sigma$ and the tangent space at that point i.e
\begin{equation} Diff_{F}(\Sigma)=\left\{ \phi \in Diff(\Sigma)| \phi(\infty)=\infty,\phi_{*}(\infty)=Id|_{T_{\infty}\Sigma} \right\}. \end{equation}
This ensures that the action of this group is free and proper when $\Sigma$ is connected and compact, which is true for the topology of $S^{3}$ that we fix here, and so superspace is ensured to be a manifold. The presymplectic form on the phase space of shape dynamics is given by
\begin{equation}\Omega_{SD}=\int_{\Sigma}\textrm{d}_{\delta}\pi^{ab}\wedge \textrm{d}_{\delta}\gamma_{ab}.\end{equation}
And, it's corresponding presymplectic potential is given by
\begin{equation}\Theta_{SD}=\int_{\Sigma}\pi^{ab}\textrm{d}_{\delta}\gamma_{ab}.\end{equation} 
First, we note that the two constraints of this theory are $Diff_{F}(\Sigma)$ and $Conf(\Sigma)$ equivariant moment maps, here $Conf(\Sigma)$ is the group of conformal transformations on $\Sigma$. This means that
$$\mathcal{C}(\rho):\Gamma_{SD}\rightarrow \mathfrak{conf}^{*}(\Sigma),$$
and
$$H_{a}(\xi^{a}):\Gamma_{SD}\rightarrow \mathfrak{diff}^{*}_{F}(\Sigma).$$
That is, they are maps from the phase space to the dual of their corresponding Lie Algebras. Their equivariance is shown via
\begin{equation}\iota_{\mathcal{J}_{\mathcal{C}(\rho)}}\Omega_{SD}=\textrm{d}_{\delta}\mathcal{C}(\rho),\end{equation}
and
\begin{equation}\iota_{\mathcal{G}_{H_{a}(\xi^{a})}}\Omega_{SD}=\textrm{d}_{\delta}H_{a}(\xi^{a}).\end{equation}
$\mathcal{G}_{H_{a}(\xi^{a})}$ has the same form as was given in the previous section, and $$\mathcal{J}_{\mathcal{C}(\rho)}=\rho\gamma_{ab}\frac{\delta}{\delta \gamma_{ab}}-\rho\pi^{ab}\frac{\delta}{\delta \pi^{ab}}.$$
On individually reducing the phase space by the two constraints we get the reduced phase spaces $$\Gamma^{\xi}_{\textrm{red}}=(H_{a}(\xi^{a}))^{-1}(0)//Diff_{F}(\Sigma),$$ and
$$\Gamma^{\rho}_{\textrm{red}}=(\mathcal{C}(\rho))^{-1}(0)//Conf(\Sigma).$$
Now, the reduced phase space of Shape Dynamics is given by
\begin{equation}\bar{\Gamma}_{SD}\cong \Gamma^{\xi}_{\textrm{red}}\cap\Gamma^{\rho}_{\textrm{red}} \cong \tilde{\Gamma}_{SD}//Diff_{F}(\Sigma)\times Conf(\Sigma).\end{equation}
Here $\bar{\Gamma}_{SD}$ is the reduced phase space of Shape Dynamics and $\tilde{\Gamma}_{SD}$ is the constraint submanifold satisfying
\begin{equation}\langle T\tilde{\Gamma}^{\perp}_{SD}|\textrm{d}\mathcal{H}_{SD}\rangle=0.\end{equation}
We shall denote the symplectic form on the reduced phase space as $\bar{\Omega}_{SD}.$ In order to attain the equations of motion on the reduced phase space, it will be convenient to `suspend' $\bar{\Omega}_{SD}$ by adding to it the two form $-\textrm{d}_{\delta}H_{SD}\wedge\textrm{d}_{\delta}\tau$. $\tau$ here is the York time given by
$$\tau=\frac{3}{2}\langle \textrm{tr}\pi\rangle. $$
The triangle brackets denote mean w.r.t $\sqrt{\gamma}$. The equation that the suspended symplectic structure has to satisfy is given by
\begin{equation}\iota_{\mathcal{X}}\bar{\Omega}_{sus}=\iota_{\mathcal{X}}[\bar{\Omega}_{SD}-\textrm{d}_{\delta}H_{SD}\wedge\textrm{d}_{\delta}\tau]=0. \end{equation}
This gives us the suspended Hamiltonian vector field
\begin{equation}\mathcal{X}=\frac{\partial}{\partial \tau}-X_{H_{SD}},\end{equation}
and
\begin{equation}X_{H_{SD}}=\frac{\delta H_{SD}}{\delta \pi^{ab}}\frac{\delta}{\delta \gamma_{ab}}-\frac{\delta H_{SD}}{\delta \gamma_{ab}}\frac{\delta}{\delta \pi^{ab}}\end{equation}
Therefore, equation (28) implies Hamilton's equations
\[\frac{\textrm{d}}{ \textrm{d} \tau}z_{SD}^{I}=\left(\begin{array}{cc}-\frac{\delta H_{SD}}{\delta \gamma_{ab}}\\ \frac{\delta H_{SD}}{\delta \pi^{ab}}\end{array}\right).\]
Here $z_{SD}^{I}$ is a point on the Shape Dynamics physical phase space given by
\[z_{SD}^{I}=\left(\begin{array}{cc}\pi^{ab}\\ \gamma_{ab}\end{array}\right).\]
\subsection{Dimensionless, Conformally Invariant Parameterization Of Phase Space}
In order to attain a dimensionless, conformally invariant parameterization of phase space, a unimodular metric
$$\bar{\gamma}_{ab}=\gamma^{-1/3}\gamma_{ab},$$ 
and the corresponding momenta
$$\sigma_{ab}=Y^{2}_{0}\gamma^{1/3}[\pi^{ab}-\frac{1}{3}\gamma^{ab}\textrm{tr}\pi].$$
Here $Y_{0}$ is some initial value of the York time, and so the new `time' in this thoery is given by the dimensionless
$$\tau=Y/Y_{0}.$$
The Symplectic structure now satisfies
\begin{equation}\bar{\Omega}_{SD}(X_{\bar{\gamma}_{ab}(x)},X_{\sigma^{cd}(x')})=Y^{2}_{0}\delta_{\textrm{T}}\delta^{3}(x-x').\end{equation}
Here we dub $\delta_{T}=\frac{1}{2}\delta^{c}_{(a}\delta^{d}_{b)}-\frac{1}{2}\bar{\gamma}_{ab}\bar{\gamma}^{ab}$ the transverse projector.
The Hamiltonian is now given by
\begin{equation}H_{SD}=\int_{\Sigma}e^{6\tilde{\phi}[\bar{\gamma},\sigma,\tau]}.\end{equation}
Here $\tilde{\phi}[\bar{\gamma},\sigma,\tau]$ contains the factor $\textrm{ln}\gamma^{\frac{1}{12}}$.
The Hamilton's equations are now given by
\[\frac{3}{2}Y_{0}^{2}\frac{\textrm{d}}{ \textrm{d} \tau}\bar{z}_{SD}^{I}=\left(\begin{array}{cc}-\frac{\delta H_{SD}}{\delta \bar{\gamma}_{ab}}\\ \frac{\delta H_{SD}}{\delta \sigma^{ab}}\end{array}\right).\]
\subsection{Geometric Quantization}
Now we shall attempt to quantize Shape Dynamics using the procedure outlined in the previous sections, but a little more care shall be taken in this sub section regarding the structures used. We ought begin with prequantization, where, we first define the pre-quantum line bundle $(L_{SD},\mathcal{D}_{\bar{\Theta}})$ over $\bar{\Gamma}_{SD}$. Apart from the usual prequantization condition which dictates that
$$\mathcal{D}_{\bar{\Theta}}\mathcal{D}_{\bar{\Theta}}=\frac{1}{h}\bar{\Omega}_{SD},$$
We have the conditions
$$\mathcal{D}_{\bar{\Theta}}(\mathcal{G}_{H_{a}(\xi^{a})})=\pi_{L_{SD}}^{*}H_{a}(\xi^{a}),$$
and
$$\mathcal{D}_{\bar{\Theta}}(\mathcal{J}_{\mathcal{C}(\rho)})=\pi_{L_{SD}}^{*}\mathcal{C}(\rho)$$
Here, $\pi_{L_{SD}}$ is the bundle projection $\pi_{L_{SD}}:L_{SD}\rightarrow \bar{\Gamma}_{SD}$.
We see that the pre-quantum connection is given by
$$\mathcal{D}_{\bar{\Theta}}=\textrm{d}_{\delta}-\frac{1}{\hbar}\bar{\Theta}$$
The polarization $P_{SD}$ is, in general, a distribution belonging to $T\bar{\Gamma}_{SD}\otimes \mathbf{C}$. In this case, we choose the polarization to be real, i.e.
$$P_{SD}=\bar{P}_{SD}.$$ More specifically, we choose the vertical polarization (locally) given by
$$P=\textrm{span}\left\{\frac{\delta}{\delta \sigma^{ab}}\right\}.$$
Thus, 
$$\mathcal{D}_{\bar{\Theta}}(X)\psi=0;\forall X\in P_{SD}.$$
Here, $\psi$ is a section of the polarized Hermitian line bundle $L^{P}_{SD}$. All that's left is to introduce the inner product on $L^{P}_{SD}$ given by

\begin{equation}\langle \Psi_{1},\Psi_{2}\rangle=\Pi\int_{(Riem(\Sigma)/Conf(\Sigma))\times \mathbf{R}^{+}} \delta \Sigma^{ab}\Psi_{1}\Psi^{*}_{2}.\end{equation}
Here, 
$$\delta\Sigma^{ab}=\epsilon^{aba_{1}b_{1}a_{2}b_{2}}\delta\bar{\gamma}_{a_{1}b_{1}}\delta\bar{\gamma}_{a_{2}b_{2}},$$
where $\epsilon^{aba_{1}b_{1}a_{2}b_{2}}$ is nowhere vanishing on the conformal superspace of Shape Dynamics.
We shall now derive the Schrodinger equation for the quantum evolution of Shape Dynamics. To begin, we first redefine the classical flow of $H_{SD}$ with respect to York time, and so it will be given by
$$f^{\tau}_{H_{SD}}=\sum_{n=0}^{\infty}\frac{\tau^{n}}{n!}X^{n}_{H_{SD}}.$$
From the Hamilton equations derived in the previous section, we know that 
$$\frac{3}{2}Y^{2}_{0}\frac{\partial}{\partial \tau}z^{I}_{SD}=X_{H_{SD}}z^{I}_{SD}.$$
Thus we can write the flow of the Hamiltonian as
$$f^{\tau}_{H_{SD}}=\sum_{n=0}^{\infty}\frac{\tau^{n}}{n!}\left(\frac{3}{2}Y^{2}_{0}\right)^{n}\frac{\partial^{n}}{\partial\tau^{n}}.$$
For any function $F$ on $\bar{\Gamma}_{SD}$, the quantum operator corresponding to it is given by
$$\hat{F}\Psi=\mathcal{Q}(F)\Psi=-i\hbar\frac{\partial}{\partial \tau}\hat{f}^{\tau}_{F}\Psi|_{\tau=0}.$$
Thus, for the Hamiltonian, we have
$$\hat{H}_{SD}\Psi=\mathcal{Q}(H_{SD})\Psi=-i\frac{3\hbar}{2}Y^{2}_{0}\frac{\partial}{\partial \tau}\Psi, $$
Which is but the time dependent Schrodinger equation for this theory.
Firstly, we see that due to the presence of the York time, we can associate the flow of the Hamiltonian with a much simpler one.
 On hindsight, it is worthy of note that Shape Dynamics solves the problem identified in the previous section with the quantization of geometrodynamics, as in this theory, there is a true Hamiltonian, which generated deterministic evolution with respect to the York time, so here, the problem of re-foliation invariance no longer persists as Shape Dynamics is known to be equivalent to ADM gravity in CMC gauge (although the notion of space-time foliation has no meaning in Shape dynamics itself). We shall now attempt to construct an evolution operator for this theory. First, the action of the quantum flow of $H_{SD}$ is given by
\begin{equation}\hat{f}^{\tau}_{H_{SD}}\Psi=\Psi_{\tau}e^{-\frac{2i}{3\hbar}Y^{-2}_{0}\int_{\tau_{0}}^{\tau}\int_{\Sigma}[\bar{\Theta}(X_{H_{SD}})-H_{SD}]\textrm{d}\tau }.\end{equation}
And so
$$\delta\Psi=\int_{\Sigma}-\frac{2i}{3\hbar}Y^{-2}_{0}H_{SD}\textrm{d}\tau$$
With this, one may construct the $\tau$ ordered unitary evolution operator, formally written as
\begin{equation}U(\tau,\tau_{0}):=Te^{-\frac{2i}{3\hbar}Y^{-2}_{0}\int_{\Sigma}\int_{\tau_{0}}^{\tau}H_{SD}\textrm{d}\tau}.\end{equation}
Here, $T$ denotes time ordering. This operator is viable since, classically, the true Shape Dynamics Hamiltonian is real and gauge invariant.
In order to connect this to equation (32) one need only note that the flows satisfy the Chapman--Kolmogorov law
$$\hat{f}^{\tau_{1}}_{H_{SD}}\circ\hat{f}^{\tau_{2}}_{H_{SD}}=\hat{f}^{\tau_{1}+\tau_{2}}_{H_{SD}},$$
Then, if one exploited the fact that
$$-i\hbar\frac{3}{2}Y^{2}_{0}\frac{\partial}{\partial \tau}\hat{f}^{\tau}_{H_{SD}}=H_{SD}\hat{f}^{\tau}_{H_{SD}},$$
to construct a formal solution for $\hat{f}^{\tau}_{H_{SD}}$, then it would be found that
$$\hat{f}^{\tau}_{H_{SD}}=U(0,\tau).$$
Similarly, the quantization of the metric is given by the operator and it's corresponding action on the wave function
\begin{equation}\mathcal{Q}(\bar{\gamma}_{ab})\Psi=\hat{\bar{\gamma}}_{ab}\Psi=\bar{\gamma}_{ab}\Psi.\end{equation}
Thus the metric acts via multiplication. Now, the conjugate TT momentum is quantized as
\begin{equation}\mathcal{Q}(\sigma^{ab})\Psi=\hat{\sigma}^{ab}\Psi=Y^{2}_{0}\delta_{T}\frac{\delta}{\delta \bar{\gamma}_{ab}}\Psi.\end{equation}
Usually, one would attempt to try and quantize the R.H.S of the Scrodinger equation on attaining these operators, but due to its implicit dependence on the metric and the momenta, it shall indeed be difficult to do so for the general case for Shape Dynamics. This problem can be tackled presently in the asymptotically flat case (see \cite{hg1}), the quantization of which shall be the subject of future work. 
   
\section{Concluding Remarks}
We have shown that Shape Dynamics is more susceptible being geometrically quantized than ADM gravity primarily due the attractive features of it's dynamics. The quantization we present here is but formal, none the less, it is intended to give the reader at least a glimpse of what the true theory of quantum gravity should look like.  Also, we have chosen the strategy reduce, then quantize, so we do not have the constraints at the quantum level, which would be the case if we did this the other way around, as both the constraints in this theory are linear in the momenta and are equivariant moment maps, quantization would commute with reduction.
\section{Acknowledgments}
The author would like to sincerely thank Henrique Gomes, Julian Barbour, Tim Koslowski and Flavio Mercati for interesting discussions regarding Shape Dynamics and for insightful comments regarding this work. This work was carried out at the \textit{Centre for Fundamental Research and Creative Education}, Bangalore, India. The author would like to acknowledge the Directors (and his mentors) Dr B.S Ramachandra and Ms. Pratiti B R for facilitating an atmosphere of free scientific inquiry so conducive to creativity. The author would also like to thank his fellow researchers Magnona H Shastry, Madhavan Venkatesh, Karthik T Vasu and Arvind Dudi.

\end{document}